\UseRawInputEncoding
\documentclass[10pt]{iopart}
\expandafter\let\csname equation*\endcsname\relax
\expandafter\let\csname endequation*\endcsname\relax
\providecommand{\keywords}[1]{\textbf{\textit{Keywords---}} #1}
\usepackage[utf8]{inputenc}
\usepackage[T1]{fontenc}
\usepackage{gensymb}
\usepackage{chemformula}
\usepackage{cite}
\usepackage{upgreek}
\begin{document}
\title{Bismuth surfactant-enhanced III-As epitaxy on GaAs(111)A}

\author{Ahmed M Hassanen$^1$\footnote{Present address:
Department of Engineering Science, University of Oxford, Parks Road, Oxford, OX1 3PJ, United
Kingdom}, Jesus Herranz$^2$\footnote{Present address:
Microsoft Quantum Materials Lab, Kanalvej 7, 2800 Kongens Lyngby, Denmark}, 
Lutz Geelhaar$^2$ and Ryan B Lewis$^{1,2}$}

\address{$^1$ Department of Engineering Physics, McMaster University, Hamilton, L8S 4L7, Canada}
\address{$^2$ Paul-Drude-Institut für Festkörperelektronik, Leibniz-Institut im Forschungsverbund Berlin e. V., Hausvogteiplatz 5–7, 10117 Berlin, Germany}
\ead{rlewis@mcmaster.ca}

\begin{abstract}
Quantum dot (QD) growth on high ($c_{3v}$) symmetry GaAs\{111\} surfaces holds promise for efficient entangled photon sources.
Unfortunately, homoepitaxy on GaAs\{111\} surfaces suffers from surface roughness/defects and InAs deposition does not natively support Stranski–Krastanov (SK) QD growth. Surfactants have been identified as effective tools to alter the epitaxial growth process of III-V materials, however, their use remains unexplored on GaAs\{111\}. Here, we investigate Bi as a surfactant in III-As molecular beam epitaxy (MBE) on GaAs(111)A substrates, demonstrating that Bi can eliminate surface defects/hillocks in GaAs and (Al,Ga)As layers, yielding atomically-smooth hillock-free surfaces with RMS roughness values as low as 0.13 nm. Increasing Bi fluxes are found to result in smoother surfaces and Bi is observed to increase adatom diffusion. The Bi surfactant is also shown to trigger a morphological transition in InAs/GaAs(111)A films, directing the 2D InAs layer to rearrange into 3D nanostructures, which are promising candidates for high-symmetry quantum dots. The desorption activation energy ($U_{Des}$) of Bi on GaAs(111)A was measured by reflection high energy electron diffraction (RHEED), yielding $U_{Des}$ = 1.7 $\pm$ 0.4 eV. These results illustrate the potential of Bi surfactants on GaAs(111)A and will help pave the way for GaAs(111)A as a platform for technological applications including quantum photonics.
\end{abstract}
\submitto{\SST}
\maketitle
\ioptwocol
\section*{Introduction}

Nanostructures such as quantum dots (QDs) grown on \{111\} surfaces have been posited as ideal entangled photon sources as a consequence of their three-fold symmetry ($c_{3v}$)\cite{Singh2009}\cite{Schliwa2009}. The production of entangled photon pairs is related to the degree of symmetry of the nanostructure, with higher symmetries minimizing fine structure splitting (FSS)\cite{Juska2013}. InAs/GaAs(100) QDs have been explored for entangled photon generation, as QD synthesis by the Stranski–Krastanov (SK) growth mode proceeds naturally on this orientation. However, (100) QDs suffer from a lack of symmetry as a consequence of anistoropic diffusion, with only a small fraction of (100) QDs within a sample being suitable for entangled photon emission\cite{Juska2013}. To overcome this, complex  post-growth processing of QDs, such
as manual selection of symmetric QDs\cite{Hafenbrak2007TriggeredK}, applying strain \cite{Seidl2006EffectDot}, and subjecting the QDs to electromagnetic fields\cite{Kowalik2005InfluenceDots}\cite{Stevenson2006APairs} have been used to mitigate the FSS issue and improve the low yield of entanglement-suitable QDs. The fabrication of symmetric InAs/GaAs\{111\} QDs is of high interest to the quantum optics community, with such quantum light sources facilitating the investigation of fundamental questions in quantum entanglement and applications in quantum information and quantum optical networks\cite{Kimble2008}. GaAs\{111\} is also a promising platform for spintronics\cite{Hernandez-Minguez2012} and topological insulators\cite{zhang2013}.

Unfortunately, epitaxy on GaAs\{111\} surfaces is complicated by several growth challenges, which have prevented development of technologies on this platform\cite{Yang1992}\cite{Esposito2017}. Firstly, the main epitaxial growth mode used for self-assembled QD synthesis\textemdash the SK growth mode\textemdash is not natively supported when depositing InAs on GaAs(111), with 2D Frank–Van der Merwe (FM) growth proceeding\cite{Hooper1993TheStudy}\cite{Joyce2004a}.Secondly, GaAs(111)A homoepitaxy is plagued by surface defects/roughness\cite{Horikoshi2007}\cite{Esposito2017} characteristic of \{111\} surfaces\cite{Einax2013ColloquiumMorphologies}. The cause of the roughness is thought to be a large positive Ehrlich-Schwöbel (ES) barrier, imposing an anistropy in the flow of adatoms across step edges by inhibiting downhill diffusion while allowing the uphill diffusion or the ``climbing up" of adatoms atop the nucleating islands\cite{Esposito2017}. Metal \{111\} surfaces have been known to exhibit large ES barriers\cite{Ferrer1999}, driving 3D growth and the creation of mounds for homoepitaxial growths. This is in contrast to GaAs(100), where the ES barrier is believed to be small and negative, encouraging net downhill diffusion and inhibiting 3D growth\cite{Tiedje2008}. For non-diffusion limited conditions, the ES barrier is the primary contributor to the rough surface morphology witnessed for GaAs(111)A\cite{Esposito2017}.

The surfactant Bi has been employed in III-V epitaxy mostly on (100) surfaces, to modify surface energetics and kinetics\cite{pillai2000}\cite{Zvonkov2000}\cite{Tixier2003}\cite{Young2005}\cite{Okamoto2010}\cite{Fan2013}. The impact of Bi during InAs/GaAs(100) QD growth has been explored\cite{Bailey2022GrowthTemperature}\cite{Alghamdi2022EffectSubstrates}, demonstrating a Bi-induced increase in QD size and optical quality\cite{Okamoto2010}\cite{Okamoto2016}, modification of QD density\cite{Fan2013}\cite{Dasika2014}, and increased dot uniformity\cite{Chen2019}. Recently, a Bi surfactant was shown to induce SK growth of InAs QDs on GaAs(110), a substrate that like GaAs(111)A does not natively support SK growth \cite{Lewis2017}\cite{Lewis2017a}\cite{Corfdir2017}\cite{Lewis2019}. However, the impact of surfactants on GaAs\{111\} has remained unexplored, and their impact on GaAs\{111\} growth problems unknown. Adding further motivation, it has been proposed that surfactants could alter ES barriers in materials with large positive barriers such as Cu(111)\cite{Hao2018}.

In this work, we explore the effect of a Bi surfactant in III-As molecular beam epitaxy on GaAs(111)A surfaces. The impact of Bi on the surface morphology of (111)A GaAs and (Al,Ga)As layers, as well as the impact on InAs/GaAs(111)A growth is studied. Bismuth is shown to completely eliminate pyramidal hillock formation in GaAs(111)A and (Al,Ga)As epitaxy, resulting in atomically smooth surfaces and, remarkably, step flow growth with terrace widths of up to 1\ $\upmu$m wide. Reflection High Energy Electron Diffraction (RHEED) experiments demonstrate that a steady-state Bi coverage is present during Bi deposition, which desorbs upon interrupting the Bi flux. The desorption energy barrier of Bi on GaAs(111)A, $U_{Des}$, was determined by modeling the RHEED data in the context of the Langmuir adsorption model, yielding $U_{Des}$ = 1.7 $\pm$ 0.4 eV. Finally, we show that exposing InAs/GaAs(111)A films to Bi can provoke the formation of InAs 3D nanostructures "on-demand", directing a 2D-to-3D transition, presenting a promising pathway for synthesizing InAs(111)A QDs. These results could be key for enabling future technological development on \{111\} III-V surfaces.

\section*{Methods}

Samples were grown using both solid- and gas-source MBE on undoped and nominally on-axis ($\pm0.5\degree$) GaAs(111)A wafers (Ga-terminated). Homoepitaxial GaAs(111)A buffer layer samples of 150 nm thickness were grown using a gas-source MBE under various Bi fluxes. The Ga and Bi fluxes were provided by solid-source effusion cells, while the As flux was provided via arsine flow cracked at 1000~$\degree$C. The substrate was heated to 350~$\degree$C, with an As flux corresponding to an equivalent GaAs growth rate of 2.25~$\upmu$m/hr being initiated at this temperature. A \ch{H_2} inductively coupled plasma (ICP) source was used to remove the native oxide layer, then the substrate was heated to 585~\degree C. Oxide desorption was confirmed by observation of a ($2~\times~2$) surface reconstruction with RHEED, which is characteristic of the GaAs(111)A surface\cite{Woolf1992}. The substrate temperature was lowered to the growth temperature of 485~\degree C, and a Ga flux was deposited at a growth rate of~0.25 $\upmu$m/s, with an As/Ga ratio of 9. Various Bi fluxes were deposited in conjunction with GaAs layer growth and the substrate was rotated during growth to ensure flux uniformity. 

(Al,Ga)As/GaAs and InAs/GaAs samples were grown by solid-source MBE, with fluxes provided by effusion cells for Ga, In, and Bi, and by a valved cracker for As$_{2}$. For these samples, the native oxide was thermally desorbed at 622~\degree C under an As$_{2}$ flux corresponding to 22 $\upmu$m/hr equivalent GaAs growth rate, which was maintained throughout the entire growth. GaAs buffer layers of 150 nm thickness or greater were grown at 485~\degree C and 0.75~$\upmu$m/hr under a Bi flux of 0.83 monolayers/s (ML/s, corresponding to the GaAs(111) monolayer atomic density). For the (Al,Ga)As samples, 300~nm thick Al$_{0.3}$Ga$_{0.7}$As layers were deposited at 1.07~$\upmu$m/hr under various Bi fluxes and T$_{sub}$~=~485 \degree C. For the InAs layer sample, 1.4~ML of InAs(111) (corresponding to 1.6~ML on GaAs(100)) was deposited on 150 nm thick GaAs buffer layers with a growth rate of 0.087~ML/s at T$_{sub}$ = 485~\degree C in the absence of Bi, resulting in the formation of a 2-dimensional InAs layer. For the InAs 3D island sample, a 2D InAs layer (grown as just described) was subsequently exposed to a Bi flux of 0.35~ML/s for 2~s immediately following InAs deposition. Sample surface morphology was investigated post-growth with atomic force microscopy (AFM) in tapping mode. 

\section*{Results \& Discussion}
\subsection*{Bi:GaAs(111)A MBE}

\begin{figure*}[ht]
\centering
\includegraphics[width=1\textwidth]{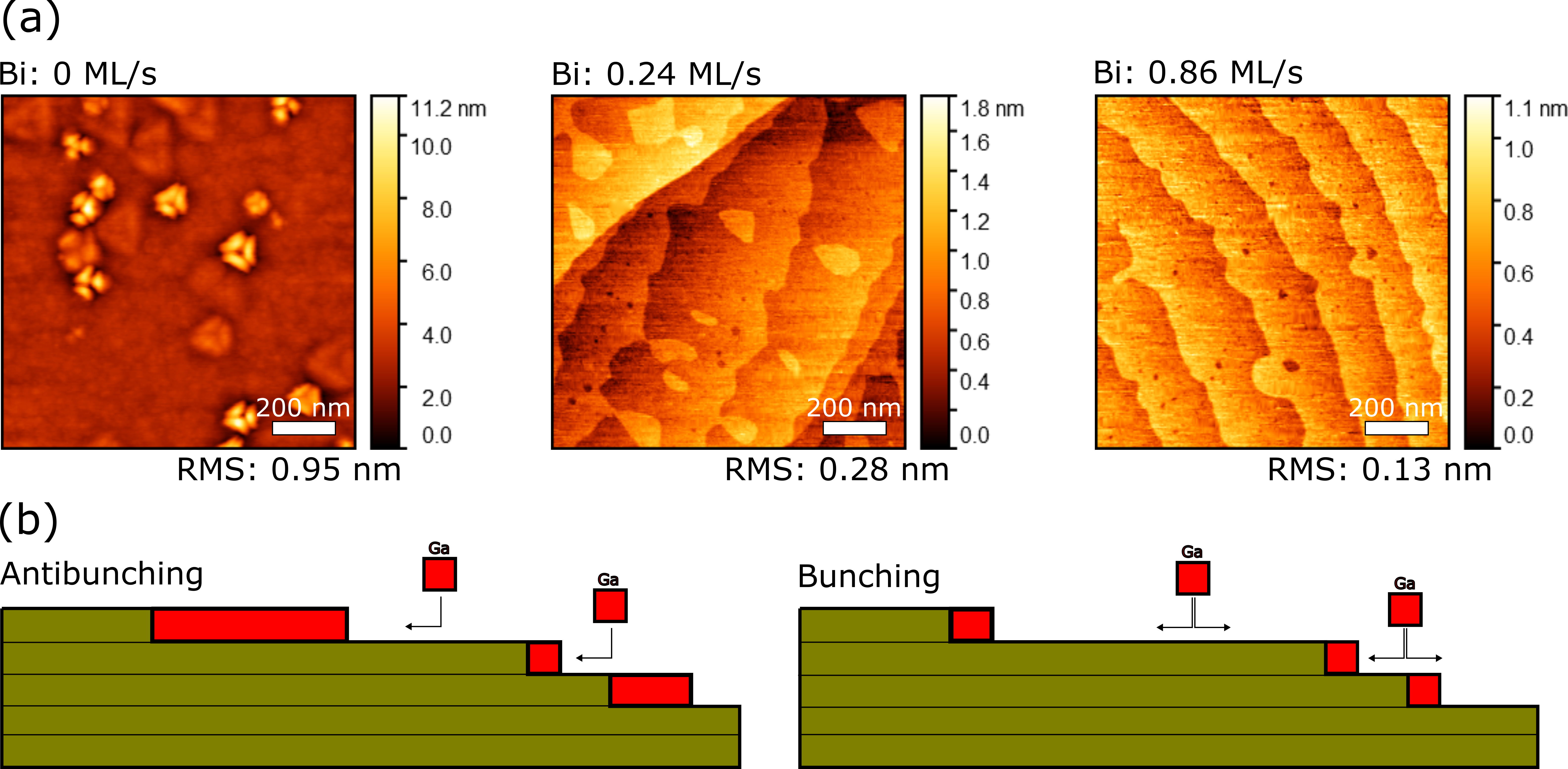}
\caption{(a) AFM topographs of GaAs(111)A layers deposited under various Bi fluxes as indicated in the figure. While growth without Bi results in surface defects, the presence of Bi results in atomically smooth surfaces, including step-flow growth for the 0.86 ML/s Bi sample. (b) Illustration of how a tendency for uphill adatom adsorption leads to step anti-bunching (step width equalization, left) while step bunching results when downhill adatom adsorption equals or exceeds uphill adsorption (right). Red bars illustrate incorporation of adatoms at step edges.}
\label{fig1}
\end{figure*}

AFM scans of GaAs(111)A buffer layers grown under various Bi fluxes are shown in Fig. \ref{fig1}a. In the absence of Bi, pyramidal dendritic defects and hillocks (height 9-12 nm and diameter about 100 nm) are present on the surface. The RMS roughness for this sample is 0.95 nm. These characteristics are consistent with previous observations in the literature\cite{Esposito2017} and highlight the challenges for technological development on GaAs(111)A. In striking contrast, the addition of a Bi flux during growth results in a radically different GaAs surface morphology. For growth with 0.24 ML/s and 0.86 ML/s Bi fluxes, atomically-smooth surfaces are observed with atomic terraces and monolayer steps. The RMS surface roughness is measured to be 0.28 nm and 0.13 nm, respectively, for these scans. For the 0.24 ML/s Bi flux growth, monolayer islands are visible on the terraces, with the terraces themselves exhibiting some step-bunching. For the sample grown under a Bi flux of 0.86 ML/s, the surface is even smoother, lacking even monolayer islands and indicating that the growth proceeded by step-flow. This is remarkable, as the wafer is nominally on-axis and the terrace widths in the topograph are several hundred nanometers wide. Furthermore, the terrace widths are evenly spaced, indicating step-antibunching during growth. We note the presence of monolayer-deep holes in the terraces, which is discussed below.

These results indicate that step-flow growth occurs despite the substrate being nominally on-axis. No substantial hillocks or defects can be found on the surface for the 0.24 and 0.86~ML/s Bi growths, which is in sharp contrast to the non-Bi growth. The surface morphology of the 0.24~ML/s Bi sample suggests that the growth is at the morphological transition between island nucleation (layer-by-layer growth) and step-flow growth. This suggests that the adatom diffusion length for this sample is similar to the terrace width (arising from the local substrate offcut), corresponding to an adatom diffusion length of 250-500~nm. This is somewhat larger than the Ga diffusion lengths quoted in literature for GaAs(111)A MBE growth, which range from 100 nm to several hundred nanometres\cite{Leys2000}\cite{Sibirev2009}\cite{Tuktamyshev2021}. Furthermore, for the sample grown under 0.86~ML/s Bi, the complete absence of any islands on the surface indicates that the diffusion length of Ga for this sample is significantly greater than the terrace width of up to 300~nm. These results suggest that the presence of surface Bi may increase Ga adatom diffusion. We note that for III-V growth on GaAs(100), there has been considerable debate about the impact of surfactant Bi on adatom diffusion \cite{Tixier2003}\cite{Okamoto2010}\cite{Fan2013}\cite{Dasika2014}\cite{Okamoto2016}\cite{Chen2019}. As mentioned above, the formation of hillocks on GaAs(111)A is believed to be a consequence of the ES barrier, resulting in uphill diffusion being favored over downhill, driving mound formation. It appears that the Bi surfactant eliminates this uphill diffusion tendency.

The even-spacing of the terraces for the  0.86~ML/s Bi sample suggests an effective diffusion length significantly greater than the terrace widths at this high Bi flux. Offcut surfaces with large ES barriers, such as some \{111\} surfaces, have been shown to support step-bunching due to the anisotropy in diffusion kinetics imposed by the barrier\cite{sato2001}\cite{Zauska-Kotur2021}. Interestingly, the ES barrier, which has been posited to be the root of the roughening and hillock formation problem in GaAs\{111\} homoepitaxial growth\cite{Esposito2017} could be provoking atomic step equalization. While it has been suggested that the mounds forming on GaAs(111)A homoepitaxial layers are fueled by the large ES barrier favoring net diffusion of adatoms uphill\textemdash driving 3D growth\textemdash that is not the case here. The presence of step equalization/antibunching indicates that adatom incorporation favors attachment to uphill step edges. If this were not the case\textemdash if downhill attachment was equally or more favored\textemdash step bunching would result. The reason for this is depicted in Fig. \ref{fig1}b. The rate at which atoms impinge on a terrace is proportional to the terrace width. Uphill attachment reduces terrace width, while downhill attachment increases it. Consequently, if uphill attachment is favored, large terraces will shrink faster than smaller ones, and so the terrace widths will equalize\textemdash antibunching. In contrast, if downhill attachment is favored, large terraces grow faster than smaller ones, leading to step bunching. We speculate that the presence of Bi atoms at step edges could be driving the incorporation at uphill step edges, while possibly enhancing downhill diffusion of adatoms. This suggests that Bi may act to lock adatoms at uphill step edges, with adatoms preferring to incorporate rather than diffuse upwards. 

Intriguingly, the GaAs layers grown with Bi exhibit monolayer holes in the terraces, with this effect more pronounced for the sample grown with the highest Bi flux. We speculate that these holes could be the result of Bi having incorporated into the surface layer (possibly on Ga sites) during growth, subsequently desorbing from the surface after the growth was stopped. The lack of Bi accumulation on the surface\textemdash despite the large flux incident during growth\textemdash indicates that a steady state coverage of Bi was present during growth, desorbing after the Bi flux was interrupted at the end of growth. This conclusion is supported by the observation of a $(1 \times 1)$ RHEED periodicity during Bi deposition, transitioning to a  $(2 \times 2)$ RHEED pattern after Bi was interrupted. Below, this RHEED transition is used to estimate the desorption energy of Bi from the surface.

\begin{figure*}[ht]
	\centering
	\includegraphics[width=0.6\textwidth]{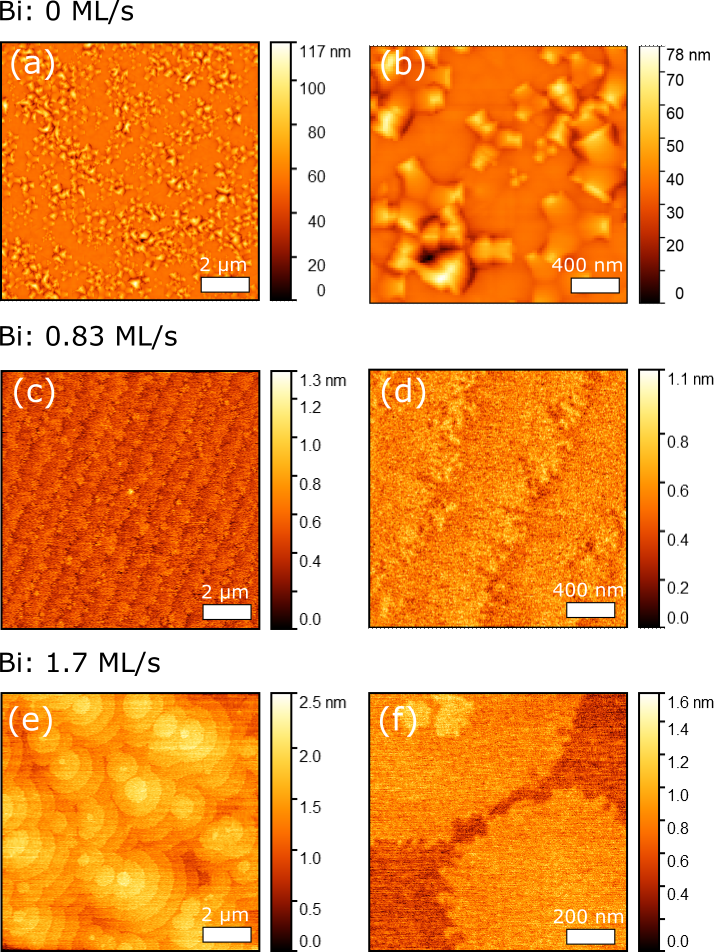}
	\caption{AFM topographs of Al$_{0.3}$Ga$_{0.7}$As/GaAs(111)A layers deposited under various Bi fluxes as indicated in the figure. (a--b) In the absence of Bi, large surface defects are present. (c--d) The presence of a Bi flux of 0.83~ML/s results in atomically smooth surfaces with notable rough step edges. (e--f) A high Bi flux of 1.7~ML/s results in a bizzarly terraced surface of regular step widths.}
	\label{fig2}
\end{figure*}
 
 \subsection*{Bi:(Al,Ga)As(111)A MBE}

In addition to the ability to grow smooth GaAs layers, high-quality (Al,Ga)As growth is required for optoelectronic device development on GaAs(111)A. Figure \ref{fig2} shows AFM images of Al$_{0.3}$Ga$_{0.7}$As layers grown under various Bi fluxes by solid-source MBE. In the absence of Bi (c.f. Fig. \ref{fig2}a--b), Al$_{0.3}$Ga$_{0.7}$As results in rough surfaces with a high density of dendritic surface defects, with an RMS roughness of 9.6 nm and an average defect height of around 20 nm. However, the addition of a 0.83~ML/s Bi flux (c.f. Fig. \ref{fig2}c--d) results in an atomically-smooth surface (RMS roughness of 0.125 nm). As for the GaAs sample grown with nearly the same Bi flux, the surface is characteristic of step-flow growth, despite the large terrace widths of about 1 $\upmu$m in these regions. In contrast to GaAs growth however, the presence of Al results in rough step edges. We speculate that this could be related to Bi desorption from the surface incorporation layer after growth, with the lower adatom mobility of Al compared to Ga making the (Al,Ga)As layers less able to rearrange after the growth is interrupted and Bi desorbs. Increasing the Bi flux to 1.7 ML/s results in a bizzare surface morphology, consisting of islands of a few monolayers in height, with a consistent step-width over the entire image, which is not the result of a local offcut. We propose that this strange morphology can be explained in terms of differing Al and Ga adatom mobilities during growth. We speculate that the Al diffusion length during growth is comparable to the terrace width, with the Ga diffusion length being much greater. In this scenario, the terrace width of islands is limited by the Al diffusion length. However, the much more mobile Ga adatoms drive step antibunching by favoring uphill attachment. The result is the smoothest possible surface as limited by the Al adatom diffusivity.
\begin{figure*}[ht]
	\centering
	\includegraphics[width=1\textwidth]{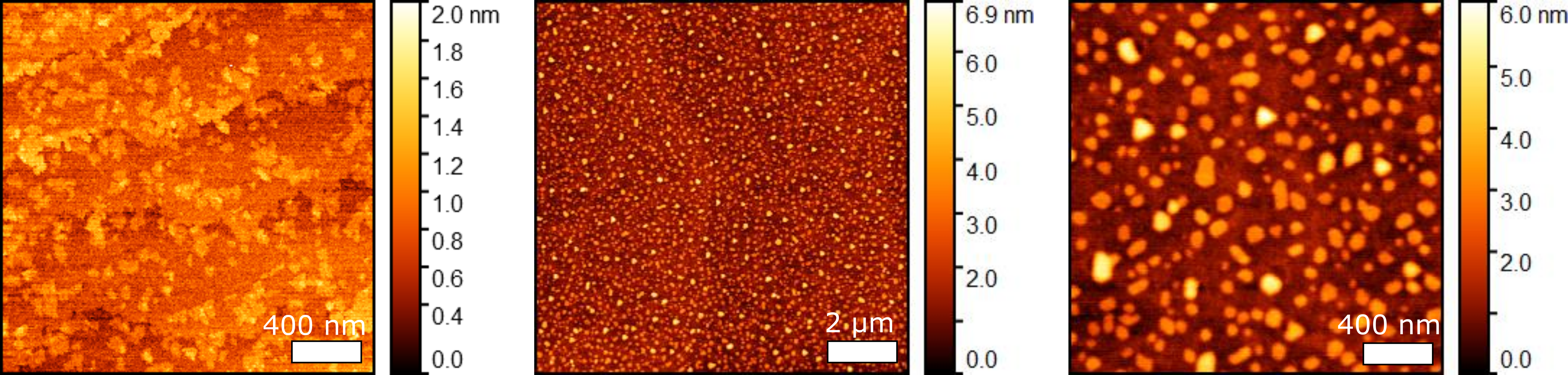}
	\caption{(a) AFM topographs of 1.4 ML thick InAs layer deposited on ultra-smooth GaAs(111)A. (b--c) After exposing a 1.4 ML InAs layer to A Bi flux of 0.35 ML/s Bi for 2 s, 3D islands are present on the surface.}
	\label{fig3}
\end{figure*}

\subsection*{InAs/GaAs(111)A growth}

The ability to realize ultra-smooth GaAs(111)A enables the investigation of InAs deposition on this surface and the impact of a Bi flux on the InAs. The surface topography of a 1.4 ML InAs layer deposited (in the absence of Bi) on ultrasmooth GaAs(111)A is shown in Fig. \ref{fig3}a. This surface exhibits monolayer islands of average size 7 nm with an average area of 240 nm$^2$ overlaid on 3--4 $\upmu$m wide terraces. This morphology is consistent with a large adatom surface diffusion for the underlying GaAs buffer growth with Bi, followed by a smaller diffusion length during InAs deposition without Bi. Clearly, the InAs deposition proceeded by a 2D growth mode here. We note that previous investigations of InAs deposition on GaAs\{111\} surfaces found the growth to proceed by a 2D growth mode, in contrast to growth on GaAs(100), for which InAs growth proceeds by the SK growth mode\cite{Joyce1997}\cite{yamaguchi1997}\cite{Ito2016}. The effect of exposing a 1.4 ML InAs/GaAs(111)A layer to Bi immediately following InAs deposition is shown in Figs. \ref{fig3}b--c. The presence of Bi promotes the formation of 3D islands on the surface. This result is consistent with our previous findings for InAs/GaAs(110) layers, where Bi was shown to provoke a morphological transition in the strained InAs layer by modifying surface energies, directing the InAs layer to spontaneously rearrange "on-demand" to form 3D islands\cite{Lewis2017}\cite{Lewis2019}. This preliminary result could pave the way to realizing high-symmetry optically-active quantum dots on GaAs(111)A, which are prospective for entangled photon emitters.

\begin{figure*}[ht]
	\centering
	\includegraphics[width=1\textwidth]{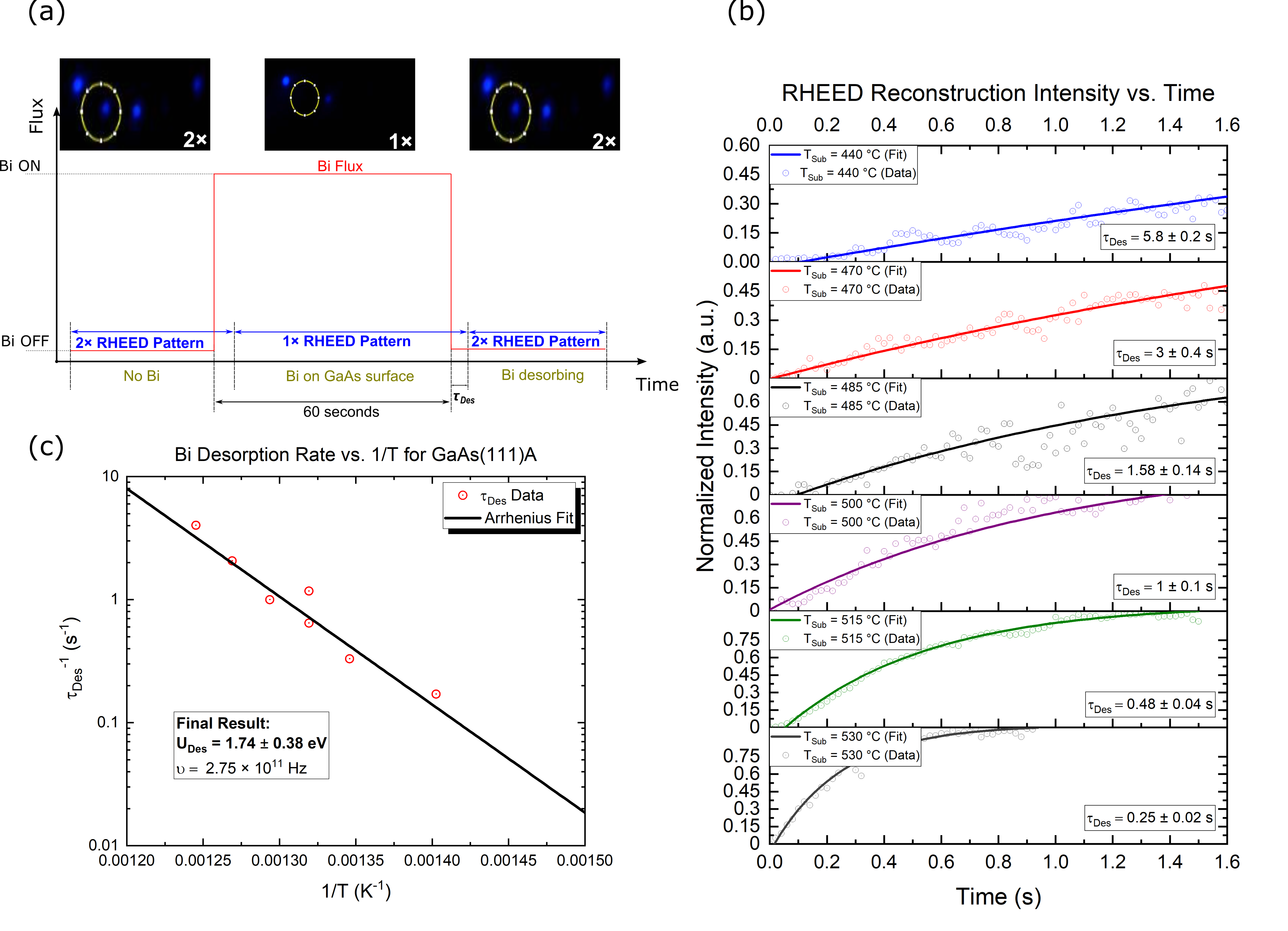}
	\caption{RHEED measurement of Bi desorption on GaAs(111)A. (a) Illustration of the flux sequence used in the experiment along with corresponding RHEED images. Before Bi deposition, a $2 \times 2$ pattern is observed, which switches to a ($1 \times 1$) pattern during Bi deposition. Upon interruption of the Bi flux, the ($2 \times 2$) pattern returns after a short delay. (b) RHEED spot intensity (points) and model fits (lines) plotted as a function of time for the reappearing RHEED streak at different substrate temperatures. The circled $2\times$ spot in (a) was used to generate the plots. The desorption times extracted from the model fit, $\tau_{Des}$ are given on each plot. (c) Arrhenius plot of the desorption rate vs. inverse temperature. The best fit line corresponds to $U_{Des}= 1.74\ eV$. $R^2 = 0.977$.}
	\label{figRHEED}
\end{figure*}
 \subsection*{RHEED Investigation of Bi Desorption}

 The samples described above were grown with large Bi fluxes up to 1.7 ML/s, which were incident on the substrate for up to an hour, corresponding to thousands of deposited monolayers of Bi. Nevertheless, no evidence of Bi surface accumulation was observed, indicating that the Bi surface coverage during deposition was in a steady state, with excess Bi thermally desorbing from the surface under these conditions. We note that similar experiments carried out at lower substrate temperatures (not shown) did result in accumulation of large amounts of surface Bi. An important parameter describing the behaviour of Bi on GaAs(111)A is the desorption energy barrier $U_{Des}$, and the desorption time for Bi on this surface $\tau_{Des}$. To measure $U_{Des}$, RHEED desorption experiments were carried out on ultra-smooth GaAs(111)A static surfaces, held at various temperatures under continuous As deposition. For the experiment, a Bi flux of 0.24~ML/s was deposited for 60 s and subsequently interrupted, as illustrated in Fig. \ref{figRHEED}a. A RHEED pattern video was recorded during the experiment, where the substrate was not rotated. Under Bi deposition, a (1 $\times$ 1) RHEED pattern is observed, which transitions back to a (2 $\times$ 2) pattern after the Bi flux is interrupted. The intensity rise of a 2$\times$ RHEED spot---assumed to correspond to the desorption time of Bi atoms on the surface---was extracted from RHEED videos by processing the individual video frames. This process was repeated at several different temperatures to determine the dependence of $\tau_{Des}$ on the temperature, enabling the extraction of the desorption activation energy from the slope of an Arrhenius fit ($\ln{(\nicefrac{1}{\tau_{Des}})}$ vs. $\nicefrac{1}{T}$). The Arrhenius relationship describing desorption rate is given as 

\begin{equation} \label{eq_lineslope}
\frac{1}{\tau_{Des}} = \nu \cdot e^{(\nicefrac{-U_{Des}}{kT})}
\end{equation}\label{eq1}where $U_{Des}$ is the desorption activation energy for Bi on GaAs(111), $T$ is the substrate temperature, $\nu$ is the attempt frequency of Bi, and $k_B$ is the Boltzmann constant. 

Figure \ref{figRHEED}b displays the intensity of the $2\times$ RHEED spots as a function of time upon interrupting Bi at various substrate temperatures $T_{Sub}$. An exponential rise is used to fit the data, which is assumed to correspond to the Bi coverage during the desorption. Assuming a Langmuir adsorption model, where the desorption time of Bi adatoms does not depend on the surface coverage, we expect the Bi coverage to decay exponentially with time. The fitting equation for the RHEED intensity $I(t)$, assumed to be proportional to the fraction of uncovered surface, is therefore

\begin{equation} \label{eq_lineslope}
I(t) = I_{Max} (1-e^{-\nicefrac{(t-x)}{\tau_{Des}}})
\end{equation}\label{eq3}
where $I_{Max}$ is the saturation RHEED intensity (steady-state after long times), and $(x)$ is a parameter used to align the rise model with the data (accounting for variations in the exact time that the Bi flux was interrupted). The measured RHEED intensity, model fits and the extracted $\tau_{Des}$ rise time values and uncertainties from the fitted curves are presented in Fig. \ref{figRHEED}b. The $R^2$ parameters signifying goodness of fit for $T_{Sub}$ = 440 $\degree$C, 470 $\degree$C, 485 $\degree$C, 500 $\degree$C, 515 $\degree$C, 530 $\degree$C are 0.9986, 0.9338, 0.8815, 0.8653, 0.9906, and 0.98 respectively.

An Arrhenius plot of the extracted desorption times is shown in Fig. \ref{figRHEED}c. The slope of this plot yields a value for the desorption energy barrier for Bi on GaAs(111)A of $U_{Des}$ = 1.7 $\pm$ 0.4 eV, with $\nu$ = 2.75 $\times$ $10^{11}$ Hz.

For the formation of a stable monolayer as per the Langmuir model, adsorption is limited to a single monolayer, with a finite value for $U_{Des}$. Therefore, it is assumed that the adsorbent adatoms can stick to the surface, but not to each other, limiting adsorption to one monolayer formation and preventing droplet formation. This is consistent with the fact that bulk Bi accumulation does not occur at the investigated conditions. We note that our measured desorption energy of Bi on GaAs(111)A is similar to the Bi self-desorption energy\cite{Herzberg1950MolecularMolecules}\cite{Young2005}. However, the large uncertainty in our $U_{Des}$  value prevents a precise comparison of values, and as noted above, Bi accumulation does occur at lower temperatures. Finally, we cannot exclude the possibility that the activation energy measured is related to surface reconstruction itself, and not Bi desorption directly.


\section*{Conclusion}

In summary, Bi surfactant action was studied in III-As MBE on GaAs(111)A. Bi acts as a surfactant for GaAs(111)A growth, inducing atomically-smooth surfaces (RMS roughness as low as 0.13 nm) and a morphological transition from island growth to step-flow growth on nominally on-axis substrates, with the effect increasing as the Bi flux is increased. Step antibunching is observed and the Ga adatom diffusion length increases as a result of Bi. We propose that Bi counteracts the Ehrlich-Schwöbel (ES) barrier responsible for 3D growth by increasing adatom adherence to step-edges and favoring uphill attachment. The Bi surfactant also profoundly improves (Al,Ga)As/GaAs(111)A growth, which has important implications for optoelectronics device development on GaAs(111)A. Furthermore, Bi can drive 3D nanostructure formation in 2D  InAs/GaAs(111)A layers, presenting a novel approach to realize (In,Ga)As QDs on GaAs\{111\}. A Bi desorption energy barrier of 1.7 $\pm$ 0.4 eV on GaAs(111)A was measured by RHEED. These findings increase understanding of this previously unstudied system of Bi on GaAs(111)A and pave the way for optoelectronic device development and QD growth on the GaAs(111)A surface, opening up avenues for QD-based quantum light sources and technological exploitation of the previously troublesome GaAs\{111\} platform.

\section*{Acknowledgments} 
We are grateful to S. Tavakoli, M. Höricke, C. Stemmler and McMaster University’s Centre for Emerging Device Technologies for MBE technical support. We graciously acknowledge financial support from the Natural Sciences and Engineering Research Council of Canada under project [RGPIN-2020-05721]. RBL acknowledges funding from the Alexander von Humboldt Foundation.

\section*{References}
\bibliography{paper}

\begin{thebibliography}{10}

\bibitem{Singh2009}
Ranber Singh and Gabriel Bester.
\newblock {Nanowire Quantum Dots as an Ideal Source of Entangled Photon Pairs}.
\newblock {\em Physical Review Letters}, 103(6):063601, 8 2009.

\bibitem{Schliwa2009}
Andrei Schliwa, Momme Winkelnkemper, Anatol Lochmann, Erik Stock, and Dieter
  Bimberg.
\newblock {In(Ga)As/GaAs quantum dots grown on a (111) surface as ideal sources
  of entangled photon pairs}.
\newblock {\em Physical Review B - Condensed Matter and Materials Physics},
  80(16), 2009.

\bibitem{Juska2013}
Gediminas Juska, Valeria Dimastrodonato, Lorenzo~O. Mereni, Agnieszka
  Gocalinska, and Emanuele Pelucchi.
\newblock {Towards quantum-dot arrays of entangled photon emitters}.
\newblock {\em Nature Photonics}, 7(7):527--531, 2013.

\bibitem{Hafenbrak2007TriggeredK}
R.~Hafenbrak, S.~M. Ulrich, P.~Michler, L.~Wang, A.~Rastelli, and O.~G.
  Schmidt.
\newblock {Triggered polarization-entangled photon pairs from a single quantum
  dot up to 30 K}.
\newblock {\em New Journal of Physics}, 9(9):315--315, 9 2007.

\bibitem{Seidl2006EffectDot}
Stefan Seidl, Martin Kroner, Alexander H{\"{o}}gele, Khaled Karrai, Richard~J.
  Warburton, Antonio Badolato, and Pierre~M. Petroff.
\newblock {Effect of uniaxial stress on excitons in a self-assembled quantum
  dot}.
\newblock {\em Applied Physics Letters}, 88(20):203113, 5 2006.

\bibitem{Kowalik2005InfluenceDots}
K.~Kowalik, O.~Krebs, A.~Lema{\^{i}}tre, S.~Laurent, P.~Senellart, P.~Voisin,
  and J.~A. Gaj.
\newblock {Influence of an in-plane electric field on exciton fine structure in
  InAs-GaAs self-assembled quantum dots}.
\newblock {\em Applied Physics Letters}, 86(4):041907, 1 2005.

\bibitem{Stevenson2006APairs}
R.~M. Stevenson, R.~J. Young, P.~Atkinson, K.~Cooper, D.~A. Ritchie, and A.~J.
  Shields.
\newblock {A semiconductor source of triggered entangled photon pairs}.
\newblock {\em Nature}, 439(7073):179--182, 1 2006.

\bibitem{Kimble2008}
H.~J. Kimble.
\newblock {The quantum internet}.
\newblock {\em Nature}, 453(7198):1023--1030, 2008.

\bibitem{Hernandez-Minguez2012}
A.~Hern{\'{a}}ndez-M{\'{i}}nguez, K.~Biermann, R.~Hey, and P.~V. Santos.
\newblock {Electrical suppression of spin relaxation in GaAs(111)b quantum
  wells}.
\newblock {\em Physical Review Letters}, 109(26):1--5, 2012.

\bibitem{zhang2013}
Dong Zhang, Wenkai Lou, Maosheng Miao, Shou-cheng Zhang, and Kai Chang.
\newblock {Interface-Induced Topological Insulator Transition in GaAs/Ge/GaAs
  Quantum Wells}.
\newblock {\em Physical Review Letters}, 156402(October):1--5, 2013.

\bibitem{Yang1992}
K.~Yang and L.~J. Schowalter.
\newblock {Surface reconstruction phase diagram and growth on GaAs(111)B
  substrates by molecular beam epitaxy}.
\newblock {\em Applied Physics Letters}, 60(15):1851--1853, 1992.

\bibitem{Esposito2017}
Luca Esposito, Sergio Bietti, Alexey Fedorov, Richard N{\"{o}}tzel, and Stefano
  Sanguinetti.
\newblock {Ehrlich-Schw{\"{o}}bel effect on the growth dynamics of GaAs(111)A
  surfaces}.
\newblock {\em Physical Review Materials}, 1(2):2--9, 2017.

\bibitem{Hooper1993TheStudy}
S~E Hooper, D~I Westwood, D~A Woolf, S~S Heghoyan, and R~H Williams.
\newblock {The molecular beam epitaxial growth of InAs on GaAs(111)B- and
  (100)-oriented substrates: a comparative growth study}.
\newblock {\em Semiconductor Science and Technology}, 8(6):1069--1074, 6 1993.

\bibitem{Joyce2004a}
Bruce~A. Joyce and Dimitri~D. Vvedensky.
\newblock {Self-organized growth on GaAs surfaces}, 12 2004.

\bibitem{Horikoshi2007}
Y.~Horikoshi, T.~Uehara, T.~Iwai, and I.~Yoshiba.
\newblock {Area selective growth of GaAs by migration-enhanced epitaxy}.
\newblock {\em physica status solidi (b)}, 244(8):2697--2706, 8 2007.

\bibitem{Einax2013ColloquiumMorphologies}
Mario Einax, Wolfgang Dieterich, and Philipp Maass.
\newblock {Colloquium : Cluster growth on surfaces: Densities, size
  distributions, and morphologies}.
\newblock {\em Reviews of Modern Physics}, 85(3):921--939, 7 2013.

\bibitem{Ferrer1999}
S~Ferrer, M~A Ni{\~{n}}o, J~E Prieto, and J~Ferr.
\newblock {Epitaxial growth of metals with high Ehrlich – Schwoebel barriers
  and the effect of surfactants}.
\newblock {\em Applied Physics A Materials Science and Processing},
  557:553--557, 1999.

\bibitem{Tiedje2008}
T.~Tiedje and A.~Ballestad.
\newblock {Atomistic basis for continuum growth equation: Description of
  morphological evolution of GaAs during molecular beam epitaxy}.
\newblock {\em Thin Solid Films}, 516(12):3705--3728, 2008.

\bibitem{pillai2000}
M~R Pillai, Seong-soo Kim, S~T Ho, S~A Barnett, and Seong-soo Kim.
\newblock {Growth of In x Ga 1 - x As / GaAs heterostructures using Bi as a
  surfactant}.
\newblock {\em Journal of Vacuum Science {\&} Technology B}, (18):1232, 2000.

\bibitem{Zvonkov2000}
B.~N. Zvonkov, I.~A. Karpovich, N.~V. Baidus, D.~O. Filatov, S.~V. Morozov, and
  Yu~Yu Gushina.
\newblock {Surfactant effect of bismuth in the MOVPE growth of the InAs quantum
  dots on GaAs}.
\newblock {\em Nanotechnology}, 11(4):221--226, 2000.

\bibitem{Tixier2003}
S.~Tixier, M.~Adamcyk, E.~C. Young, J.~H. Schmid, and T.~Tiedje.
\newblock {Surfactant enhanced growth of GaNAs and InGaNAs using bismuth}.
\newblock {\em Journal of Crystal Growth}, 251(1-4):449--454, 2003.

\bibitem{Young2005}
E.~C. Young, S.~Tixier, and T.~Tiedje.
\newblock {Bismuth surfactant growth of the dilute nitride GaNxAs 1-x}.
\newblock {\em Journal of Crystal Growth}, 279(3-4):316--320, 2005.

\bibitem{Okamoto2010}
Hiroshi Okamoto, Takehiko Tawara, Hideki Gotoh, Hidehiko Kamada, and Tetsuomi
  Sogawa.
\newblock {Growth and characterization of telecommunication-wavelength quantum
  dots using Bi as a surfactant}.
\newblock {\em Japanese Journal of Applied Physics}, 49(6 PART 2), 2010.

\bibitem{Fan2013}
Dongsheng Fan, Zhaoquan Zeng, Vitaliy~G. Dorogan, Yusuke Hirono, Chen Li,
  Yuriy~I. Mazur, Shui~Qing Yu, Shane~R. Johnson, Zhiming~M. Wang, and
  Gregory~J. Salamo.
\newblock {Bismuth surfactant mediated growth of InAs quantum dots by molecular
  beam epitaxy}.
\newblock {\em Journal of Materials Science: Materials in Electronics},
  24(5):1635--1639, 2013.

\bibitem{Bailey2022GrowthTemperature}
N.~J. Bailey, M.~R. Carr, J.~P.~R. David, and R.~D. Richards.
\newblock {Growth of InAs(Bi)/GaAs Quantum Dots under a Bismuth Surfactant at
  High and Low Temperature}.
\newblock {\em Journal of Nanomaterials}, 2022(001):1--9, 6 2022.

\bibitem{Alghamdi2022EffectSubstrates}
Haifa Alghamdi, Amra Alhassni, Sultan Alhassan, Amjad Almunyif, Alexey~V
  Klekovkin, Igor~N Trunkin, Alexander~L Vasiliev, Helder~V.A. Galeti,
  Yara~Galvão Gobato, Igor~P Kazakov, and Mohamed Henini.
\newblock {Effect of bismuth surfactant on the structural, morphological and
  optical properties of self-assembled InGaAs quantum dots grown by Molecular
  Beam Epitaxy on GaAs (001) substrates}.
\newblock {\em Journal of Alloys and Compounds}, 905(February):164015, 6 2022.

\bibitem{Okamoto2016}
Hiroshi Okamoto.
\newblock {Self-Organized Nanostructure Formation of III-V and IV
  Semiconductors with Bismuth}.
\newblock {\em Journal of Advances in Nanomaterials}, 1(2):82--94, 2016.

\bibitem{Dasika2014}
Vaishno~D. Dasika, E.~M. Krivoy, H.~P. Nair, S.~J. Maddox, K.~W. Park, D.~Jung,
  M.~L. Lee, E.~T. Yu, and S.~R. Bank.
\newblock {Increased InAs quantum dot size and density using bismuth as a
  surfactant}.
\newblock {\em Applied Physics Letters}, 105(25), 2014.

\bibitem{Chen2019}
X.~Y. Chen, Y.~Gu, Y.~J. Ma, S.~M. Chen, M.~C. Tang, Y.~Y. Zhang, X.~Z. Yu,
  P.~Wang, J.~Zhang, J.~Wu, H.~Y. Liu, and Y.~G. Zhang.
\newblock {Growth mechanisms for InAs/GaAs QDs with and without Bi
  surfactants}.
\newblock {\em Materials Research Express}, 6(1), 2019.

\bibitem{Lewis2017}
Ryan~B. Lewis, Pierre Corfdir, Hong Li, Jesús Herranz, Carsten Pf{\"{u}}ller,
  Oliver Brandt, and Lutz Geelhaar.
\newblock {Quantum Dot Self-Assembly Driven by a Surfactant-Induced
  Morphological Instability}.
\newblock {\em Physical Review Letters}, 119(8):1--6, 2017.

\bibitem{Lewis2017a}
Ryan~B. Lewis, Pierre Corfdir, Jesús Herranz, Hanno K{\"{u}}pers, Uwe Jahn,
  Oliver Brandt, and Lutz Geelhaar.
\newblock {Self-Assembly of InAs Nanostructures on the Sidewalls of GaAs
  Nanowires Directed by a Bi Surfactant}.
\newblock {\em Nano Letters}, 17(7):4255--4260, 2017.

\bibitem{Corfdir2017}
Pierre Corfdir, Ryan~B. Lewis, Lutz Geelhaar, and Oliver Brandt.
\newblock {Fine structure of excitons in InAs quantum dots on GaAs(110) planar
  layers and nanowire facets}.
\newblock {\em Physical Review B}, 96(4):1--7, 2017.

\bibitem{Lewis2019}
Ryan~B. Lewis, Achim Trampert, Esperanza Luna, Jesús Herranz, Carsten
  Pf{\"{u}}ller, and Lutz Geelhaar.
\newblock {Bismuth-surfactant-induced growth and structure of InAs/GaAs(110)
  quantum dots}.
\newblock {\em Semiconductor Science and Technology}, 34(10), 2019.

\bibitem{Hao2018}
Jialei Hao and Lixin Zhang.
\newblock {Surface Science Strongly reduced Ehrlich – Schwoebel barriers at
  the Cu ( 111 ) stepped surface with In and Pb surfactants}.
\newblock {\em Surface Science}, 667(September 2017):13--16, 2018.

\bibitem{Woolf1992}
D.~A. Woolf, Z.~Sobiesierski, D.~I. Westwood, and R.~H. Williams.
\newblock {The molecular beam epitaxial growth of GaAs/GaAs(111)B: Doping and
  growth temperature studies}.
\newblock {\em Journal of Applied Physics}, 71(10):4908--4915, 1992.

\bibitem{Leys2000}
M~R Leys.
\newblock {Fundamental growth kinetics in MOMBE / CBE , MBE and MOVPE}.
\newblock {\em Journal of Crystal Growth}, (209):225--231, 2000.

\bibitem{Sibirev2009}
Nickolay~Vladimirovich Sibirev, V.~G. Duborvskii, E.~B. Arshanskii, G~E Cirlin,
  Y.~B. Samsonenko, and V.~M. Ustinov.
\newblock {On Diffusion Lengths of Ga Adatoms on AlAs ( 111 ) and GaAs ( 111 )
  Surfaces}.
\newblock {\em Technical Physics}, 54(4):586--589, 2009.

\bibitem{Tuktamyshev2021}
Artur Tuktamyshev, Alexey Fedorov, Sergio Bietti, Stefano Vichi, Riccardo
  Tambone, Shiro Tsukamoto, and Stefano Sanguinetti.
\newblock {Nucleation of Ga droplets self ‑ assembly on GaAs ( 111 ) A
  substrates}.
\newblock {\em Scientific Reports}, (111):1--11, 2021.

\bibitem{sato2001}
Masahide Sato and Makio Uwaha.
\newblock {Growth law of step bunches induced by the Ehrlich-Schwoebel effect
  in growth}.
\newblock {\em Surface Science}, 493(1-3):494--498, 2001.

\bibitem{Zauska-Kotur2021}
Magdalena Za{\l}uska-Kotur, Hristina Popova, and Vesselin Tonchev.
\newblock {Step Bunches, Nanowires and Other Vicinal
  “Creatures”—Ehrlich–Schwoebel Effect by Cellular Automata}.
\newblock {\em Crystals}, 11(9):1135, 9 2021.

\bibitem{Joyce1997}
B.~A. Joyce, J.~L. Sudijono, J.~G. Belk, H.~Yamaguchi, X.~M. Zhang, H.~T.
  Dobbs, A.~Zangwill, D.~D. Vvedensky, and T.~S. Jones.
\newblock {A scanning tunneling microscopy-reflection high energy electron
  diffraction-rate equation study of the molecular beam epitaxial growth of
  InAs on GaAs(001), (110) and (111)A - Quantum dots and two-dimensional
  modes}.
\newblock {\em Japanese Journal of Applied Physics, Part 1: Regular Papers and
  Short Notes and Review Papers}, 36(6 SUPPL. B):4111--4117, 1997.

\bibitem{yamaguchi1997}
Hiroshi Yamaguchi, J.~G. Belk, X.~M. Zhang, J.~L. Sudijono, M.~R. Fahy, T.~S.
  Jones, D.~W. Pashley, and Bruce~A. Joyce.
\newblock {Atomic-scale imaging of strain relaxation via misfit dislocations in
  highly mismatched semiconductor heteroepitaxy : InAs / GaAs „ 111 {\ldots}
  A}.
\newblock {\em Physical Review B}, 55(3):1337--1340, 1997.

\bibitem{Ito2016}
Tomonori Ito, Toru Akiyama, and Kohji Nakamura.
\newblock {Theoretical Investigations for Strain Relaxation and Growth Mode of
  InAs Thin layers on GaAs(111)A}.
\newblock {\em Condensed Matter}, 1(1):4, 2016.

\bibitem{Herzberg1950MolecularMolecules}
Gerhard Herzberg.
\newblock {\em {Molecular spectra and molecular structure. Vol.1: Spectra of
  diatomic molecules}}.
\newblock D. Van Nostrand Company. Inc., Princeton, NJ, 2 edition, 1950.

\end{thebibliography}
\bibliographystyle{unsrt}

\end{document}